\documentclass[final]{siamltex}
\usepackage{times}
\usepackage{graphics,epsfig,graphicx,color}
\usepackage{amsmath}
\usepackage{mathrsfs,amsfonts,amssymb}
 \usepackage{bm}
\usepackage{url}
\usepackage[breaklinks,colorlinks=true,linkcolor=blue,citecolor=red,backref=page]{hyperref}

\DeclareMathAlphabet{\itbf}{OML}{cmm}{b}{it}

\def\bq{{{\itbf q}}}

\def\br{{{\itbf r}}}

\def\by{{{\itbf y}}}
\def\bx{{{\itbf x}}}

\def\bk{{{\itbf k}}}

\def\bzeta{{\boldsymbol{\zeta}}}
\def\bxi{{\boldsymbol{\xi}}}

\def\btheta{{\boldsymbol{\theta}}}
\def\bpsi{{\boldsymbol{\psi}}}

\def\eps{{\varepsilon}}

\renewcommand{\d}{{\rm d}}

\newcommand{\RR}{\mathbb{R}}

\newcommand{\EE}{\mathbb{E}}

\newcommand{\rc}{\textcolor{black}}

\newtheorem{remark}{Remark}[section]

\begin{document}  
 
\title{Non-invasive imaging through random media}

\author{Josselin
Garnier\footnotemark[1]
and
 Knut S\O lna\footnotemark[2]
}

\maketitle

\renewcommand{\thefootnote}{\fnsymbol{footnote}}

\footnotetext[1]{Centre de Math\'ematiques Appliqu\'ees, Ecole Polytechnique,
91128 Palaiseau Cedex, France (josselin.garnier@polytechnique.edu)}
\footnotetext[2]{Department of Mathematics, 
University of California, Irvine CA 92697
(ksolna@math.uci.edu)}

\renewcommand{\thefootnote}{\arabic{footnote}}

\begin{abstract}
When waves propagate through a strongly scattering medium   the energy
is transferred to the incoherent wave part by scattering. The wave intensity then 
forms  a random speckle pattern  seemingly without much useful
information.  However, a number of recent physical  experiments  show how
one can extract useful  information from this speckle pattern. 
Here we present the mathematical analysis that explains the quite stunning
performance of such a  scheme for speckle imaging. 
Our analysis  identifies a scaling regime
where the scheme  works well.
This regime is the white-noise paraxial regime, which
leads to the It\^o-Schr\"odinger model for the wave 
amplitude. 
The  results presented in this paper conform  with the 
sophisticated physical intuition that has motivated these schemes, but
give a more detailed characterization of the performance. 
 The analysis gives a   description of (i) the information that 
can be extracted and with what resolution (ii)  the statistical stability
or signal-to-noise ratio with which the information can be extracted.
   \end{abstract}



\pagestyle{myheadings}
\thispagestyle{plain}

\section{Introduction}

In many contexts of wave propagation the medium is complex and vary on multiple
scales. This  is  the case for instance for the turbulent atmosphere, the fluctuating ocean,
 the complex earth's crust,  and biological tissue.
Due to small-scale scattering the coherent or mean part of the wave
can completely vanish, all energy is transferred to the incoherent
 part, and the intensity of the transmitted wave then has the form of a speckle pattern. 
 In such situations imaging techniques based on using the coherent 
 wave field  indeed  fail.
   However, in the statistics 
 of the complex speckle pattern there may be hidden
 useful information.
 
 A  physical experiment showing how the speckle pattern
 can be exploited for imaging through a complex medium is presented in \cite{katz14} 
 in the context of optics. A time-harmonic
 point source transmits a wave which  travels through a complex 
 medium. The transmitted intensity is recorded by a camera. It has the form
 of a speckle pattern because of scattering.
The autocorrelation of this  speckle pattern is typically 
 a sharply peaked  function with the peak radius being of the order of
 the correlation radius  of the speckle pattern.
  Moreover, this autocorrelation function is statistically stable (i.e., it does
  not depend on the realization of the speckle pattern, but only on its statistics)
  if the averaging (when computing the autocorrelation) takes place over many 
independent speckle spots, that is, if the camera  is large compared 
  to the speckle size. 
Consider next another time-harmonic
 point source in the neighborhood of the first point source.
 It generates another speckle pattern on the camera.
 However, by the memory effect \cite{freund88}, this speckle pattern is essentially
a shifted version of the first speckle pattern, if the two source points are not too far from each other.  
 Assume finally that  we have a  spatially incoherent extended source, 
 then  only relatively small interference takes place between the different speckle patterns, 
 and the camera image is simply the
superposition of the different speckle patterns generated by the points in the source support. 
Within the memory effect these speckle patterns are shifted 
versions of approximately the same speckle pattern.  
Then in fact the speckle pattern associated with the incoherent extended source
is the convolution of the source pattern with the speckle pattern generated by one point source.
The autocorrelation function of the speckle pattern
is then the autocorrelation function of the speckle pattern generated by one point source
 convolved with the autocorrelation function of the source pattern. 
 Since the autocorrelation function of the speckle pattern generated by one point source is sharply peaked,
this gives more or less the  autocorrelation function of the source pattern.
As a result, this can be used to unravel the source pattern via a phase  retrieval step.
 The paper \cite{katz14} elegantly presents this imaging scheme and physical results that show that this process works very well.      
  
The physical experiment that more directly motivates the analysis and modeling
in this paper is the one presented in \cite{bertolotti}.  Here, the set-up slightly differs from
the one above.
A time-harmonic laser beam propagates through a scattering medium and illuminates an object placed behind the medium.
This object is fluorescent and reemits light at a different frequency. The fluorescent light propagates through the random 
and the total intensity is collected by a bucket detector.  
We remark that measuring the total, or spatially
integrated, intensity, gives a robust scheme. This is because the total intensity 
emitted by the fluorescent  object does not change during propagation through the medium which is 
  assumed  to be lossless.
A sequence of measurements is generated by varying the 
incident angle for the probing incoming beam.
Based  on these measurements the authors in \cite{bertolotti,bertolotti2,Guo}
find
that it is possible  to estimate the shape
of the (2D) fluorescent object. 
 
 We remark that the physical phenomena behind the two experiments in  \cite{bertolotti} and \cite{katz14}  are
 analogous. In both experiments the memory effect and rapid
 decorrelation of the speckle pattern are exploited so that the autocorrelation
 of the observations essentially becomes the autocorrelation of the object to be imaged. 
 Forming the autocorrelation is very efficient in mitigating the effects
 of the random medium.  The experiments in    \cite{bertolotti} and \cite{katz14}  
 are however different in that in  \cite{katz14} a ``one shot'' image is taken 
 while the  approach in   \cite{bertolotti}  requires scanning over 
 incident angle.
     
The principle of speckle imaging is related to and can be seen 
as a further generalization of techniques associated with refocusing problems
 \cite{edrei16,mosk12,rotter,vellekoop10,vellekoop07}.
In the refocusing problem, phase conjugation or time reversal of waves lead to a sharp focusing at the original source point 
and this mechanism is the same as the one giving a stable sharp empirical
covariance function for the speckle pattern in the above experiments in the situation with one
point source only.  
The experiment we model in this paper moreover  bears similarities with
ghost imaging where also a bucket  
detector is used \cite{garnier_ghost,li10,shapiro12}.  It is different from ghost imaging   in that 
the covariance of the bucket measurements themselves are  computed, for different incident angles, 
rather than with respect to the measurements of a reference multi-element sensor array that does not see the object.

The concept  of refocusing and speckle imaging   has recently received a lot of attention in
the physical literature   and many experiments have been carried
in the vein described above, see for instance also
\cite{katz14,katz12,popoff10,vellekoop08} and the review  in \cite{rotter}. 
Here we present a novel mathematical
analysis that gives quantitative answers to questions about the performance
of such schemes  based on modeling of the propagation phenomenon from first
principles. 
    
Associated with the physical experiments described above there are indeed several 
fundamental and important questions to answer from the mathematical viewpoint:
(i)
under which scaling regime can we expect the above procedure
to work well;
(ii)
what is the resolution we can expect in the computed image,
that is, what is the degree of blurring in the image;
(iii)
what is the signal-to-noise ratio or relative amount of noise 
in the image.
Below we will give precise  mathematical answer to these questions 
when we consider propagation in the so called paraxial scintillation regime
corresponding to high-frequency waves, 
long propagation distances, and a beam radius larger than the correlation radius of the  medium. 
In particular regarding question (i) we remark that it is important
for the scheme that we have a strong memory effect.  That
is a  source shift should give  essentially  only a shift in  the 
generated speckle pattern.   Regarding question (ii) we find that 
the fundamental resolution limit of the procedure is limited 
by the characteristic speckle size. Moreover, regarding question
(iii) we find that a good signal-to-noise ratio essentially 
requires  sampling  over a broad cone of incident angles for the incoming 
beam.  These results are derived in Section \ref{sec:scint3} and we summarize  
them  in Section \ref{sec:summ} giving a quantitative characterization of the 
performance of the method.   
     
 The outline of the paper is as follows.  In Section \ref{sec:setup}
 we describe the experiment that we want to model and which 
 is motivated by the physical experiment in \cite{bertolotti}.
 The  quantity measured in  the experiment is the 
 total wave intensity transmitted or reflected by the object to be imaged as a function of source angle.
The empirical covariance function (as a function of the source angle) 
of these measurements is described in Section \ref{sec:par}. 
  In order to be able to analyze this quantity of interest
 we must specify the regime of propagation and how the 
 speckle statistics can be described in this context.
 The main scaling configuration that we consider is the paraxial or beam propagation
 regime  described in Section \ref{sec:paraxial}. 
   In particular we model the fine-scale medium fluctuations as a random field.  
We give the It\^o-Schr\"odinger equation in Section \ref{sec:paraxial1} which
 was derived from the wave equation in \cite{garniers1}. This is a forward
 or Markov approximation that describes how wave energy is transferred
 from the coherent to the incoherent part as the wave propagates through
 the scattering medium.  Scattering produces 
 a term   in the It\^o-Schr\"odinger equation that  involves a   Brownian field  
 whose lateral   statistics is inherited from the statistics of the random medium. 
 It is furthermore important to note that
 the intensity is a quadratic quantity of the field and its covariance
 is, therefore, a fourth-order moment of the field, that we need to characterize.
 Based on the It\^o-Schr\"odinger equation we can  use It\^o's calculus 
 to identify transport equations for all the moments of the wave field, in particular the fourth-order moments.  
 We can readily solve the equations for the first-order and second moments, 
 giving the mean and the covariance of the wave field, but
 in the general paraxial regime we cannot solve the fourth-order moment equation
 explicitly.  However, in the paraxial scintillation regime,  corresponding to  the physical context 
of scintillation that we want to capture,  we can in fact do so explicitly \cite{garniers4}.
The scintillation regime corresponds to a secondary scaling limit associated with a
 beam whose radius is larger than the correlation radius of the medium.
We describe this regime and the associated fourth-order moment characterization
in Section \ref{sec:scint}.  
The fourth-order moment results presented in Section 
\ref{sec:scint2} are new and capture the source configuration of interest with a varying 
incident angle for the probing field.   Then in Section \ref{sec:scint3} we  derive
the result that describes how the object can be imaged from the measured
total intensity covariance over incident angle.  
In Section \ref{sec:summ} we then summarize the  main result.
In particular we show how our mathematical results can be given a 
physical interpretation. We also discuss how a strong memory effect,
which indeed is important for the procedure to work well, 
can be interpreted in terms of the  {shower curtain effect}. 
 Finally we  present concluding  remarks in Section~\ref{sec:final}.

\section{ Physical configurations enabling  speckle imaging}
\label{sec:setup}
\begin{figure}
\begin{center}
\begin{tabular}{c}
\includegraphics[width=8.0cm]{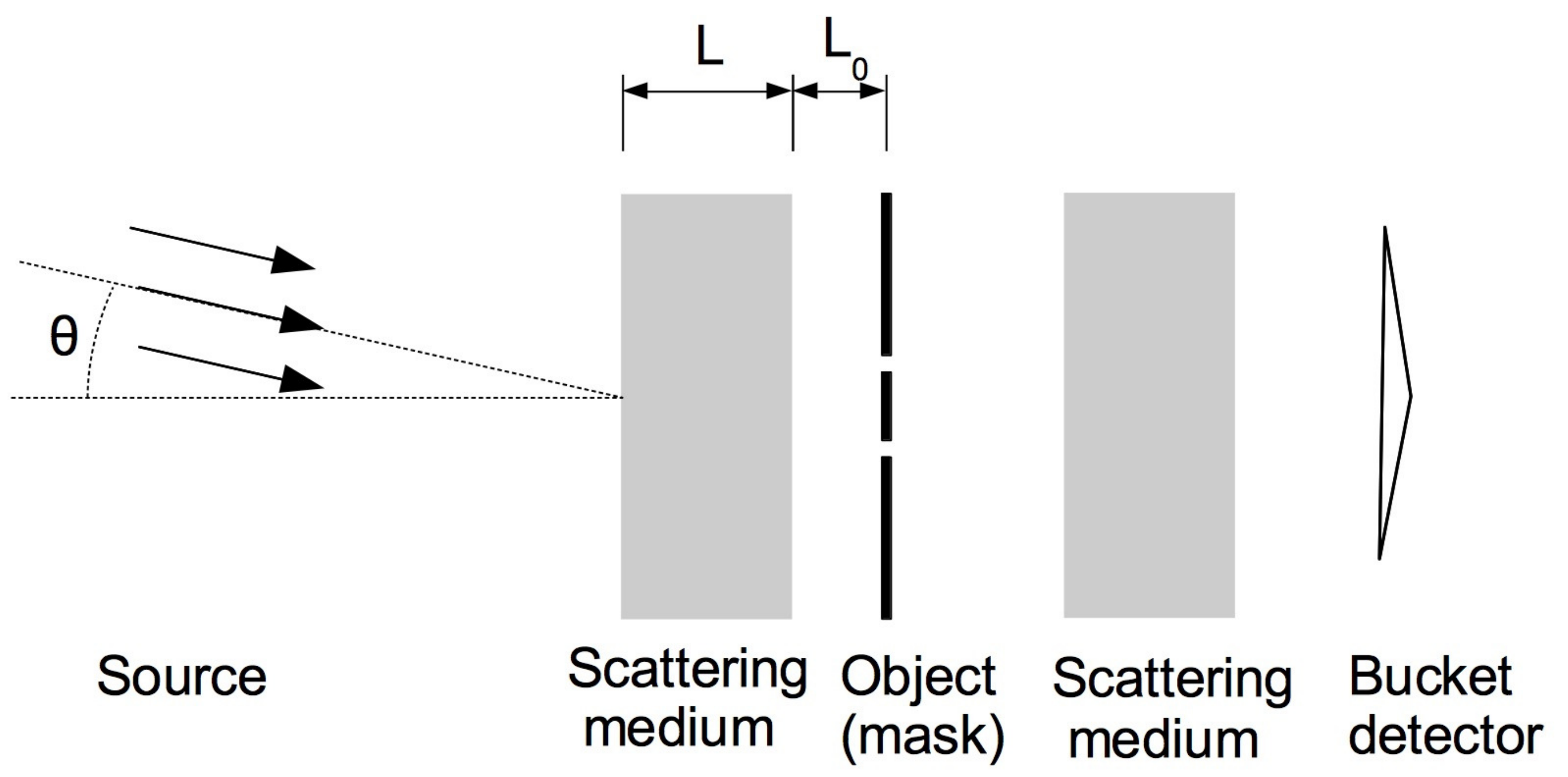}  
\end{tabular}
\end{center}
\caption{The  experimental imaging set-up. The source is time-harmonic.
A mask is placed behind a scattering medium. 
The total intensity of the light that goes through the mask is collected by a bucket detector (it may go through 
a second scattering medium or not, as this does not affect the total transmitted intensity).
For each incident angle $\btheta$ the total transmitted intensity is measured.}
\label{fig:1} 
\end{figure}

In this section we describe the physical configuration which 
is motivated by the experiment  in \cite{bertolotti}.
We consider and study the following experiment  (see Figure \ref{fig:1}): \\
1) A  mask (for instance, a double slit) is placed behind a scattering medium.\\
2) A laser beam with incident angle $\btheta$ is shined on the scattering medium, whose transmitted light produces a speckle pattern that illuminates
the mask.\\
3) The total light ${\cal E}_\btheta$ transmitted through the mask (and through a second scattering medium or not)
is collected and measured by a bucket detector.
The experiment is repeated for the same medium, mask, and source, but with different incident angles $\btheta$.
Our goal is to show that the covariance function (in $\btheta$) of the measured transmitted light intensity ${\cal E}_\btheta$
is related to the shape of the mask, moreover, how the mask can be recovered from
the empirical covariance function of the measurements.

\begin{figure}
\begin{center}
\begin{tabular}{c}
\includegraphics[width=6.0cm]{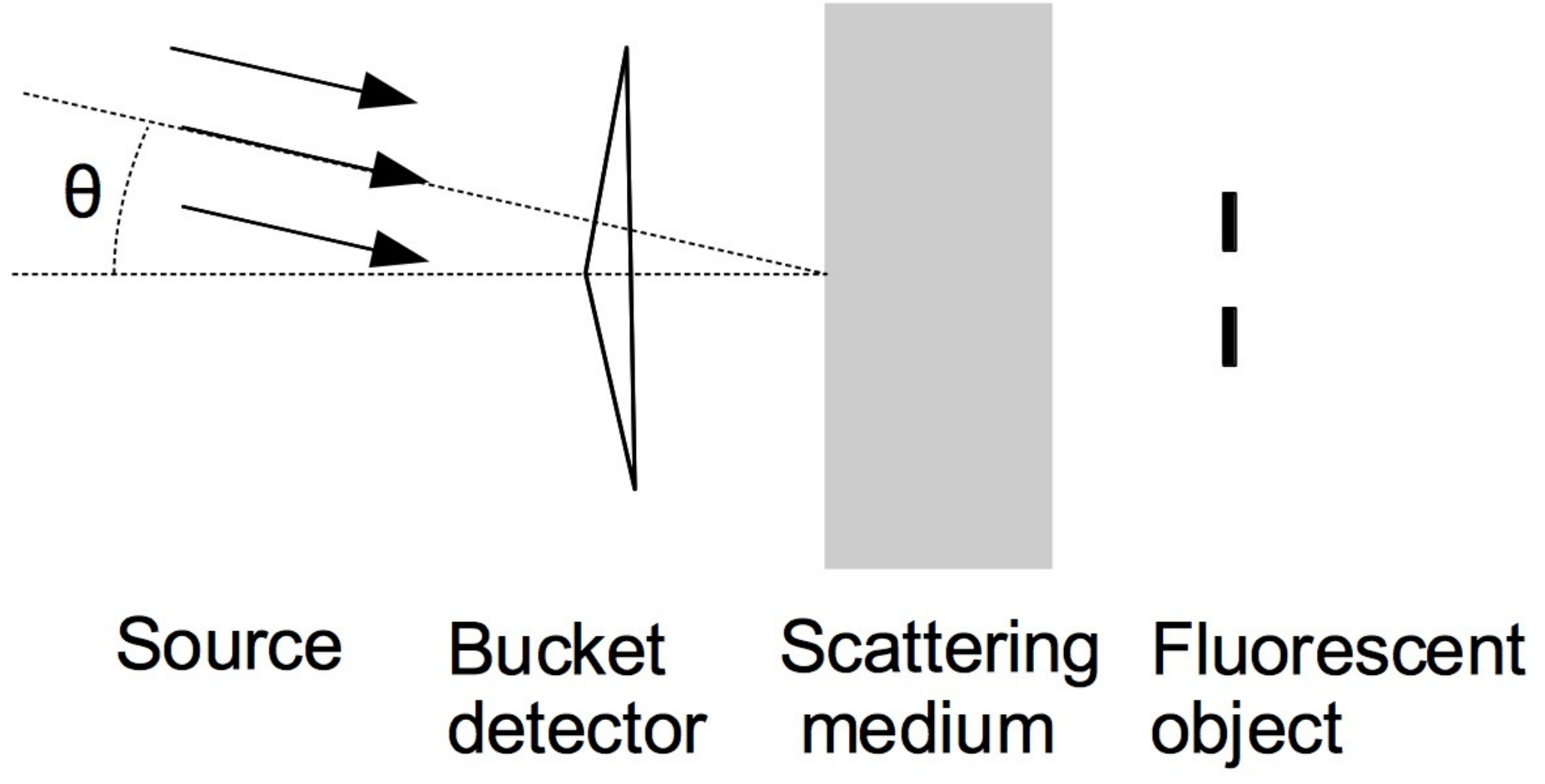}  
\end{tabular}
\end{center}
\caption{The second experimental imaging set-up \cite{bertolotti}. The source is time-harmonic.
The object is fluorescent and located behind a scattering medium.
It reemits light at a frequency that is different from the one of the source.
For each incident angle $\btheta$ the total fluorescent light is measured.}
\label{fig:2} 
\end{figure}
 
As indicated above, this experiment has some 
resemblance with other speckle intensity correlation imaging methods 
\cite{garnier_ghost,katz14,katz12,li10,popoff10,shapiro12}.
 More precisely it is equivalent to (and motivated by) a recent experiment reported in \cite{bertolotti}  (see Figure \ref{fig:2}): \\
1) A  fluorescent object is placed behind a scattering medium.\\
2) A laser beam  with incident angle $\btheta$ is shined on the scattering medium, whose transmitted light produces a speckle pattern that illuminates
the object.\\
3) The fluorescent light emitted by the object is transmitted through the same scattering medium and the total amount ${\cal E}_\btheta$
of transmitted fluorescence is collected using a band-pass filter (whose central frequency is the fluorescent frequency,
which is different from the one of the incoming laser beam) and measured by a bucket detector.\\
The experiment is repeated for the same medium, object, and source, but with different incident angles $\btheta$.
In practice, the change in the incident angle can be achieved by rotating mirrors as described in \cite{bertolotti}
or by spatial light modulators.
Mathematically, the expression of the transmitted fluorescence is equal to the expression of the transmitted intensity
in the previous experiment.
In \cite{bertolotti} the speckle memory effect  \cite{freund88} is invoked to 
identify  the relation between the 
covariance function of the transmitted fluorescence and the fluorescent object profile.
Here  we will carry out a detailed analysis to clarify under which circumstances one can indeed image the object
and which resolution one can anticipate.
The analysis is based on recent results on fourth-order moments for the random paraxial wave equation \cite{garniers4}.

\section{The measured intensity}
\label{sec:par}
\label{sec:intcor}%
The spatial variable is denoted by $(\bx,z)\in \RR^d \times \RR$.
The source is time-harmonic and has the form of an extended beam going along an axis that makes an angle $\btheta \in \RR^d$,
such that $|\btheta|\leq 1$ and $(\btheta,\sqrt{1-|\btheta|^2})$ is the unit-vector that determines the direction of the beam axis.
Its frequency is $\omega$
and its wavenumber $k_o=\omega/c_o$, with $c_o$ the background velocity:
\begin{equation}
\label{eq:inc}
E_\btheta(\bx,0) = f(\bx) \exp (i k_o \btheta \cdot \bx)  .
\end{equation}
At some point we will take the limit of a plane wave $f(\bx)=1$ but we will carry out the analysis with a general beam.
The time-harmonic field in the plane of the mask is denoted by $E_\btheta ( \bx ,L+L_0) $,
where $L$ is the thickness of the scattering medium and $L_0$ is the distance from the scattering medium to the mask.
The mask is characterized by a compactly supported non-negative valued function $U(\bx)$,
the mask indicator function.
The intensity measured by the detector is 
\begin{equation}
\label{def:Etheta}
{\cal E}_\btheta = \int_{\RR^d} |E_\btheta(\bx,L+L_0) |^2 U(\bx) \d\bx  .
\end{equation}
The covariance function of ${\cal E}_\btheta$ is our main quantity of interest.
We will see that it allows us to reconstruct $U(\bx)$.

In the second experiment of \cite{bertolotti}, 
the object is characterized by a compactly supported non-negative valued function $U(\bx)$,
the object's fluorescence response, and the total fluorescence emitted by the object and measured by the detector is 
also equal to (\ref{def:Etheta}).
 

\section{The white-noise paraxial model}\label{sec:paraxial}
\label{sec:ito}%
The model for the time-harmonic field at the output plane of the scattering medium is
\begin{equation}
\label{eq:model1}
E_\btheta ( \bx ,L) = \int_{\RR^d} \hat{g}\big((\bx,L), (\bx',0) \big) E_\btheta ( \bx' ,0)  \d \bx' ,
\end{equation}
where $E_\btheta ( \bx ,0) $ is the incident field (\ref{eq:inc}) and $\hat{g}$ is the 
fundamental solution of the white-noise paraxial wave equation 
which we describe in the next subsections. There should be an additional factor $\exp(i k_o L)$ in (\ref{eq:model1})
but it does not play any role as we only record intensities.

The model for the time-harmonic field at the plane of the mask is
\begin{equation}
\label{eq:model1b}
E_\btheta ( \bx ,L+L_0) = \int_{\RR^d} \hat{g}_0\big((\bx,L+L_0), (\bx',L) \big) E_\btheta ( \bx' ,L)  \d \bx' ,
\end{equation}
where $E_\btheta ( \bx ,L) $ is the  field (\ref{eq:model1}) and $\hat{g}_0$ is the 
fundamental solution of the homogeneous paraxial wave equation (\ref{eq:green0:chap10}).

\subsection{The random paraxial wave equation}\label{sec:paraxial1}
We consider the time-harmonic form of the scalar wave equation
\begin{equation}
(\partial_z^2+\Delta) E +  k_o^2 \big(1 + \mu(\bx,z)\big) E=0  ,
\end{equation}
where $\Delta$ is the transverse Laplacian (ie the Laplacian in $\bx$).
Here $\mu$ is a zero-mean, stationary, $d+1$-dimensional random process
with mixing properties in the $z$-direction.
The white-noise paraxial regime is a regime in which the 
typical wavelength is much smaller than the initial field radius and the correlation radius of the medium,
which are themselves much smaller than the propagation distance.
In the special high-frequency regime
\begin{equation}
\label{eq:scaleparax:chap10}
k_o \to \frac{k_o}{\delta^4} , 
\quad  \quad  
\mu(\bx,z)  \to \delta^3 \mu\big(  \frac{\bx}{\delta^2}, \frac{z}{\delta^2}\big)  ,
\end{equation}
the rescaled function $\hat{\phi}^\delta$ defined by
\begin{equation}
\label{eq:scaleparax1:chap10}
E^\delta( \bx,z)  = \exp \Big(  i \frac{k_o z}{\delta^4}  \Big) \hat{\phi}^\delta \Big( \frac{\bx}{\delta^2},z \Big) 
\end{equation}
satisfies
\begin{equation}
\label{eq:scaleparax2:chap10}
\delta^4 \partial_{z}^2  \hat{\phi}^\delta
+ 
\left(  2 i  k_o \partial_z \hat{\phi}^\delta+ \Delta \hat{\phi}^\delta + \frac{k_o^2}{\delta}  \mu\big(  \bx  , \frac{z}{\delta^2}\big)  \hat{\phi}^\delta \right)=0  .
\end{equation}
The ansatz (\ref{eq:scaleparax1:chap10}) corresponds to a plane wave 
with a slowly varying envelope.
In the regime $\delta \ll 1$, it has been shown in \cite{garniers1} that
the forward-scattering approximation and the white-noise approximation are valid,
which means that the second-order derivative in $z$ in (\ref{eq:scaleparax2:chap10}) can be neglected and 
the random potential $\frac{1}{\delta}  \mu\big(  \bx  , \frac{z}{\delta^2}\big)$ can be replaced by a white noise in $z$.
The mathematical statement is that  the function $\hat{\phi}^\delta( \bx,z)$ 
as $\delta \to 0$
converges weakly to the solution $\hat{\phi}( \bx,z)$ of the
It\^o-Schr\"odinger equation
\begin{equation}
 2 i  k_o \d_z \hat{\phi} ( \bx,z) +\Delta \hat{\phi} ( \bx,z) \d z + k_o^2  \hat{\phi}( \bx,z)\circ \d B(\bx,z) =0  ,
\end{equation}
where $B(\bx,z)$ is a Brownian field, that is a Gaussian process with mean zero and covariance function
\begin{equation}
\label{def:covgaus:chap10}
\EE\big[ B(\bx,z)B(\bx',z')\big] =  \gamma_o(\bx-\bx') \big( z \wedge z' \big),
\end{equation}
with
\begin{equation}
\label{def:gamma0:chap10}
\gamma_o(\bx)=  \int_{-\infty}^\infty  \EE[ \mu({\bf 0},0)\mu(\bx,z) ] \d z .
\end{equation}
Here the $\circ$ stands for the Stratonovich stochastic integral.

\subsection{The fundamental solution}\label{sec:paraxial2}
The fundamental solution is defined as the solution
of the It\^o-Schr\"odinger equation in $(\bx,z)$:
\begin{equation}
\label{def:greens:chap10}
 2i k_o \d_z \hat{g} + \Delta \hat{g}  \d z+ k_o^2 \hat{g} \circ \d B(\bx,z)= 0, 
\end{equation}
starting from $\hat{g}\big( (\bx,z=z_0),(\bx',z_0 ) \big) = \delta(\bx-\bx')$.
In a homogeneous medium ($B \equiv 0$) the fundamental solution is (for $z> z_0$)
\begin{equation}
\label{eq:green0:chap10}
\hat{g}_0 \big(  (\bx,z), (\bx',z_0) \big) =
\Big( \frac{ k_o}{2 i \pi (z-z_0)} \Big)^{d/2}  \exp \Big(  i \frac{k_o
      |\bx-\bx'|^2}{2  (z-z_0)} \Big)    .
\end{equation}
In a random medium, the 
first two moments of the random fundamental solution have the following expressions.

\begin{proposition}
\label{prop:parax2:chap10}%
The first order-moment of the random  fundamental solution exhibits  damping
(for $z > z_0$):
\begin{equation}
\EE \big[ \hat{g}\big( (\bx,z),(\bx',z_0) \big)   \big] 
=
\hat{g}_0\big( (\bx,z),(\bx',z_0) \big) 
 \exp \Big( -\frac{\gamma_o({\bf 0}) k_o^2 (z-z_0)}{8}  \Big)  ,
  \label{eq:mom1parax1:chap10}
\end{equation}
where $\gamma_o$ is given by (\ref{def:gamma0:chap10}).

The second order-moment of the random  fundamental solution exhibits spatial decorrelation:
\begin{align}
\nonumber
 \EE \big[ \hat{g}\big( (\bx_1,z),(\bx',z_0) \big) 
\overline{\hat{g}\big( (\bx_2,z),(\bx',z_0) \big)} \big]  =& \,
\hat{g}_0\big( (\bx_1,z),(\bx',z_0) \big) 
\overline{\hat{g}_0\big( (\bx_2,z),(\bx',z_0) \big)} \\
&\times
 \exp \Big( -  \frac{ \gamma_2(\bx_1-\bx_2) k_o^2 (z-z_0)}{4}   \Big)   , 
 \label{eq:mom2parax1:chap10}
\end{align}
where 
\begin{equation}
\gamma_2(\bx)= \int_0^1 \gamma_o({\bf 0}) -\gamma_o(\bx s) \d s .
\end{equation}
\end{proposition}

These are classical results  (see \cite[Chapter 20]{ishimaru} and \cite{garniers2})
once the the random paraxial equation has been proved to be correct, as is the case here.

The result on the first-order moment (Eq.~(\ref{eq:mom1parax1:chap10}))
allows us  to identify the scattering mean free path  
(the critical propagation distance through the random medium beyond which the coherent or mean field vanishes):
\begin{equation}
\label{def:lsca:parax:chap10}
\ell_{\rm sca} = \frac{8 }{\gamma_o({\bf 0}) k_o^2},
\end{equation}
It shows that any coherent wave imaging method 
cannot give good images if the propagation distance is larger than the scattering mean free path 
because the coherent wave components are then exponentially damped.
This is the situation we have in mind in this paper.

The result on the second-order moment (Eq.~(\ref{eq:mom2parax1:chap10}))
shows that, in the strongly scattering regime $z-z_0\gg \ell_{\rm sca}$,
when $\gamma_o$ can be expanded as
\begin{equation}
\label{eq:expandgamma0}
\gamma_o(\bx) = \gamma_o({\bf 0}) - \frac{1}{2} \bar{\gamma}_2 |\bx|^2 + o(|\bx|^2),
\end{equation}
 then the correlation radius of the field is 
$$
X_c(z-z_0) =\frac{\sqrt{12}}{\sqrt{\bar{\gamma}_2} k_o \sqrt{z-z_0}} .
$$
As seen in this formula, when the propagation distance $z-z_0$ becomes very large, then the correlation radius 
becomes of the order of the wavelength. This is the limit of validity of our results,
because when this happens, the random paraxial approximation is not valid anymore.
Therefore the paraxial distance $\ell_{\rm par}$ such that $k_o X_c(\ell_{\rm par})=1$ can be defined by 
\begin{equation}
\label{def:ltr}
\ell_{\rm par} = \frac{12}{\bar{\gamma}_2}
.
\end{equation}
In this paper we consider only propagation distances smaller than $\ell_{\rm par}$.

\begin{remark}
If the autocovariance function of the medium fluctuations $\nu$ is of the form
$\EE [\nu(\bx,z)\nu(\bx',z')] = 
\sigma_{\rm med}^2 C_{\rm med}(|\bx-\bx'|/\ell_{\rm med}, |z-z'|/\ell_{\rm med})$,
where $\ell_{\rm med}$ is the correlation radius of the medium and $\sigma^2_{\rm med}$ is the relative standard deviation
of its fluctuations, then 
$$
\ell_{\rm sca} \approx \frac{\lambda_o^2}{\sigma^2_{\rm med} \ell_{\rm med}} ,\quad \quad
\ell_{\rm par} \approx \frac{\ell_{\rm med}}{\sigma^2} .
$$
In our scaling regime $\lambda_o \ll \ell_{\rm med}$ so that $\ell_{\rm sca} \ll \ell_{\rm par}$.
\end{remark}


\subsection{The mean intensity}\label{sec:paraxial3}
In our paper the first quantity of interest is the mean intensity
\begin{equation}
\EE [ {\cal E}_\btheta] = \int \EE \big[ |E_\btheta(\bx,L+L_0) |^2 \big] U(\bx) \d\bx  .
\end{equation}
The mean Wigner transform is defined by 
\begin{equation}
{W}_\btheta ( \bx,\bxi,z) := 
\int_{\RR^d}
\exp \big( - i  \bxi \cdot \bq  \big)
\EE \left[ E_\btheta\big(  \bx+\frac{\bq}{2} ,z\big)
 \overline{E}_\btheta \big(    \bx-\frac{\bq}{2},z\big) \right]  \d \bq  ,
\end{equation}
that is, it  the angularly-resolved  mean wave energy density, and it 
satisfies the closed system
\begin{equation}
\label{systemWT2rapid}
\frac{\partial {W}_\btheta }{\partial z} + \frac{1}{k_o}
{ \bxi}\cdot \nabla_{\bx} W_\btheta 
=\frac{k_o^2}{4 (2\pi)^d}
\int_{\RR^d} \hat{\gamma}_0( \bk) \Big[ 
W_\btheta  (   \bxi - \bk   ) 
 - W_\btheta  (  \bxi ) 
\Big]  \d \bk ,
\end{equation}
starting from:
$$
W_\btheta(\bx,\bxi,0) = \int_{\RR^d} \exp \big( i (k_o \btheta \cdot \bq - \bxi \cdot \bq) \big) f \big( \bx + \frac{\bq}{2}\big) \overline{f} \big( \bx - \frac{\bq}{2}\big)  
\d \bq.
$$
By taking a Fourier transform in $\bx$ and an inverse Fourier transform in $\bxi$  
of Eq.~(\ref{systemWT2rapid}), we obtain a transport equation
that can be solved and we find
 the following integral representation for $W_\btheta$:
\begin{align}
\nonumber
W_\btheta (\bx,\bxi,L+L_0)  =& \,
\frac{1}{(2\pi)^d }\int_{\RR^d} \exp \Big(i \bzeta \cdot\big( \bx - \bxi \frac{L+L_0}{k_o}\big) 
- i \bxi\cdot \bq \Big)
\hat{W}_{\btheta} (\bzeta,\bq,0) \\
 &\times 
 \exp \Big( \frac{k_o^2}{4} \int_0^L \gamma_o \big( \bq + \bzeta \frac{z}{k_o} \big)- \gamma_o(
 {\bf 0}) \d z \Big)
 \d\bzeta \d \bq ,
\end{align}
where $\hat{W}_\btheta(\bzeta,\bq,0)$ is defined in terms of  the initial field 
$f$ as:
\begin{align}
\nonumber
\hat{W}_\btheta(\bzeta,\bq,0)  =& \, \int_{\RR^d}
\exp \big(  i (k_o \btheta \cdot \bq - \bzeta \cdot \bx ) \big)
  f\big(   \bx+\frac{\bq}{2} \big)
 \overline{f} \big(     \bx-\frac{\bq}{2}\big)    \d \bx \\
 =& \,
 \frac{1}{(2\pi)^d} e^{i k_o \btheta\cdot\bq} \int_{\RR^d}
 e^{i \bk \cdot \bq} \hat{f}\big(\bk+\frac{\bzeta}{2}\big)\overline{\hat{f}}\big(\bk-\frac{\bzeta}{2}\big)
 \d\bk  .
\end{align}
The mean intensity that illuminates the mask is therefore
\begin{align}
\nonumber
\EE \big[ E_\btheta(\bx,L+L_0)|^2\big] 
=& \,
\frac{1}{(2\pi)^d} \int_{\RR^d} e^{i \bzeta \cdot \bx} \hat{W}_\btheta\big( \bzeta , - \bzeta \frac{L+L_0}{k_o},0\big)\\
& \times \exp\Big( \frac{k_o^2}{4} \int_0^L \gamma_o(-\bzeta \frac{z+L_0}{k_o}) -\gamma_o({\bf 0}) \d z \Big) \d \bzeta   ,
\end{align}
and the mean transmitted intensity is
\begin{align}
\nonumber
\EE\big[{\cal E}_\btheta]
= &\, 
\frac{1}{(2\pi)^d} \int_{\RR^d} \overline{ \hat{U}}(\bzeta) \hat{W}_\btheta\big( \bzeta , - \bzeta \frac{L+L_0}{k_o},0\big)\\
&\times \exp\Big( \frac{k_o^2}{4} \int_0^L \gamma_o(-\bzeta \frac{z+L_0}{k_o}) -\gamma_o({\bf 0}) \d z \Big) \d\bzeta  .
\label{eq:meanfluo2}
\end{align}

\subsection{The intensity covariance function}\label{sec:paraxial4}
In our paper the main quantity of interest is the intensity covariance function  
\begin{align}
\nonumber
{\rm Cov}( {\cal E}_{\btheta},{\cal E}_{\btheta'}) =& \, \iint
 \EE \big[ |E_\btheta(\bx,L+L_0) |^2 |E_{\btheta'}(\bx',L+L_0) |^2 \big] U(\bx) U(\bx') \d\bx \d\bx'  \\
& \,  - 
\EE [ {\cal E}_\btheta]\EE [ {\cal E}_{\btheta'}]  .
\end{align}
We can write
\begin{equation}
 \EE \big[ |E_\btheta(\bx,L+L_0) |^2 |E_\btheta(\bx',L+L_0) |^2 \big] =
{\cal M}_{\btheta\btheta'} ( \bx,\bx',\bx,\bx',L+L_0) ,
\end{equation}
where the fourth-order moment 
${\cal M}_{\btheta\btheta'} ( \bx_1,\bx_2,\by_1,\by_2,z)$ 
is the solution of
\begin{align}
& \frac{\partial {\cal M}_{\btheta\btheta'} }{\partial z} = \frac{i}{2k_o}  \Big(  \Delta_{\bx_1}+\Delta_{\bx_2}
-  \Delta_{\by_1} -\Delta_{\by_2}\Big) {\cal M}_{\btheta\btheta'} + \frac{k_o^2}{4} {\cal U} \big( \bx_1,\bx_2, \by_1,\by_2,z\big)
{\cal M}   , \\
\nonumber
&  {\cal M}_{\btheta\btheta'} ( \bx_1,\bx_2,\by_1,\by_2,z=0) 
=
f (\bx_1) \overline{f (\by_1)}
f (\bx_2) \overline{f(\by_2)} \\
&\hspace*{1.8in} \times \exp\big( i k_o \btheta\cdot (\bx_1-\by_1)+ i k_o \btheta' \cdot (\bx_2-\by_2)\big),
 \end{align}
where  the generalized potential ${\cal U}$ is zero if $z \not\in [0,L]$ and 
\begin{eqnarray}
{\cal U}\big( \bx_1,\bx_2, \by_1,\by_2 ,z\big)  
=
\sum_{j,l=1}^2 \gamma_o(\bx_j-\by_l) 
- \gamma_o( \bx_1-\bx_2)
-  \gamma_o( \by_1-\by_2) -
2\gamma_o({\bf 0})  \, ,
\hspace*{0.3in}
\end{eqnarray}
if $z\in [0,L]$.

We parameterize  the four points 
$\bx_1,\bx_2,\by_1,\by_2$  in the special way:
\begin{align}
\label{eq:reliexr1}
\bx_1 = \frac{\br_1+\br_2+\bq_1+\bq_2}{2}, \quad \quad 
\by_1 = \frac{\br_1+\br_2-\bq_1-\bq_2}{2}, \\
\bx_2 = \frac{\br_1-\br_2+\bq_1-\bq_2}{2}, \quad \quad 
\by_2 = \frac{\br_1-\br_2-\bq_1+\bq_2}{2}.
\label{eq:reliexr2}
\end{align}
We denote by $\mu_{\btheta\btheta'}$ the fourth-order moment in these new variables:
\begin{equation}
\mu_{\btheta\btheta'}  (\bq_1,\bq_2,\br_1,\br_2,z) := 
{\cal M}_{\btheta\btheta'}  (
\bx_1  ,
\bx_2 ,
\by_1 ,  
\by_2 ,z
) ,
\end{equation}
with $\bx_1,\bx_2,\by_1,\by_2$ given by (\ref{eq:reliexr1}-\ref{eq:reliexr2}) in terms of $\bq_1,\bq_2,\br_1,\br_2$.
The Fourier transform (in $\bq_1$, $\bq_2$, $\br_1$, and $\br_2$) of the fourth-order moment
is defined by:
\begin{align}
\nonumber
\hat{\mu}_{\btheta\btheta'} (\bxi_1,\bxi_2,\bzeta_1,\bzeta_2,z) 
= & \, 
\iint {\mu}_{\btheta\btheta'} (\bq_1,\bq_2,\br_1,\br_2,z)  \\
& \hspace*{-0.8in}
\times
\exp  \big(- i\bq_1 \cdot \bxi_1- i\bq_2 \cdot \bxi_2- i\br_1\cdot \bzeta_1- i\br_2\cdot \bzeta_2\big) \d \bq_1\d \bq_2 
\d \br_1 \d \br_2 \label{eq:fourier} 
. 
\end{align}
It satisfies
\begin{align}
\nonumber
&
\frac{\partial \hat{\mu}_{\btheta\btheta'} }{\partial z} + \frac{i}{k_o} \big( \bxi_1\cdot \bzeta_1+   \bxi_2\cdot \bzeta_2\big) \hat{\mu}_{\btheta\btheta'}
\\
\nonumber
& =
\frac{k_o^2}{4 (2\pi)^d} {\bf 1}_{[0,L]}(z) 
\int_{\RR^d} \hat{\gamma}_0(\bk) \Big[  
 \hat{\mu}_{\btheta\btheta'}  (  \bxi_1-\bk, \bxi_2-\bk, \bzeta_1, \bzeta_2)  \\
\nonumber
& \quad  + 
 \hat{\mu}_{\btheta\btheta'}  (  \bxi_1-\bk,\bxi_2,  \bzeta_1, \bzeta_2-\bk)    
 +
 \hat{\mu}_{\btheta\btheta'}  (  \bxi_1+\bk, \bxi_2-\bk, \bzeta_1, \bzeta_2)   \\
\nonumber
&  \quad 
+ 
 \hat{\mu}_{\btheta\btheta'}  (  \bxi_1+\bk,\bxi_2, \bzeta_1,  \bzeta_2-\bk)    -
2 \hat{\mu}_{\btheta\btheta'} (\bxi_1,\bxi_2, \bzeta_1, \bzeta_2) \\
&  \quad 
-
 \hat{\mu}_{\btheta\btheta'} (  \bxi_1,\bxi_2-\bk, \bzeta_1, \bzeta_2-\bk)  
- \hat{\mu}_{\btheta\btheta'}  (  \bxi_1,\bxi_2+\bk,  \bzeta_1, \bzeta_2-\bk) 
\Big] \d \bk ,
\label{eq:fouriermom0}
\end{align}
starting from 
\begin{align}
\nonumber
& \hat{\mu}_{\btheta\btheta'}  (  \bxi_1,\bxi_2, \bzeta_1, \bzeta_2,z=0)  
=
\hat{f}\Big( \frac{\bxi_1+\bxi_2+\bzeta_1+\bzeta_2}{2} -k_o \btheta\Big)
\overline{\hat{f}}\Big( \frac{\bxi_1+\bxi_2-\bzeta_1-\bzeta_2}{2}-k_o \btheta\Big) \\
& \hspace*{0.4in} \times 
\hat{f}\Big( \frac{\bxi_1-\bxi_2+\bzeta_1-\bzeta_2}{2}-k_o \btheta'\Big)
\overline{\hat{f}}\Big( \frac{\bxi_1-\bxi_2-\bzeta_1+\bzeta_2}{2}-k_o \btheta'\Big)  .
\end{align}
The second moment of the intensity can be expressed in terms of $ \hat{\mu}$ as
\begin{align}
\nonumber
\EE \big[ {\cal E}_\btheta {\cal E}_{\btheta'} \big] 
=& \,\frac{1}{(2\pi)^{4d}} 
\iint
 \hat{\mu}_{\btheta\btheta'}  (  \bxi_1,\bxi_2, \bzeta_1, \bzeta_2,L+L_0) \\
& \times  \overline{\hat{U}(\bzeta_2+\bzeta_1)} {\hat{U}(\bzeta_2-\bzeta_1)} 
\d \bxi_1\d \bxi_2 \d \bzeta_1 \d  \bzeta_2  .
\label{eq:covfluo2}
\end{align}
No closed-form expression of the fourth-order moment of the field or of the second-order moment of the intensity is available,
but it is possible to get explicit expressions in  the scintillation regime,
which corresponds to the case where the correlation radius of the medium is smaller  than the incident field radius.

\section{The scintillation regime}\label{sec:scint}
\label{sec:scin}%
The scintillation regime is valid if the (transverse) correlation radius of the Brownian field 
(ie the transverse correlation radius of the medium fluctuations)
is smaller than the incident field radius.
We moreover assume that the standard deviation of the Brownian field is small and that the propagation distance is large.
If the correlation radius of the medium is our reference length, this means that 
in this regime the covariance function $\gamma_o^\eps$ is of the form:
\begin{equation}
\label{sca:sci}
\gamma_o^\eps(\bx)= \eps \gamma_o (\bx) , 
\end{equation}
the incident field radius is of order $1/\eps$ and the angle is small, i.e. of order one, so that the incident field is of the form 
\begin{equation}
\label{def:feps}
E^\eps_\btheta(\bx,0) = f ( \eps \bx ) \exp( i k_o \btheta \cdot\bx) ,  
\end{equation}
the mask is small, i.e. its radius is of order one, so that the mask indicator function is of the form
\begin{equation}
U^\eps (\bx) = U(\bx),
\end{equation}
and the propagation distance is of order of $1/\eps$~:
\begin{equation}
L^\eps = \frac{L}{\eps}, \quad \quad L_0^\eps=\frac{L_0}{\eps} .
\end{equation}
Here $\eps$ is a small dimensionless parameter and we will study the limit $\eps \to 0$.
Note that this problem was analyzed in \cite{garniers4} when $\btheta={\bf 0}$ and 
$f$ has a Gaussian profile. The forthcoming proposition \ref{prop:sci1} is an extension of this original result.

\subsection{The fourth-order moments}\label{sec:scint1}
Let us denote the rescaled function
\begin{equation}
\label{eq:renormhatM2}
\tilde{\mu}^\eps_{\btheta\btheta'} (\bxi_1,\bxi_2,\bzeta_1,\bzeta_2,z) := 
\hat{\mu}_{\btheta\btheta'} \Big(\bxi_1,\bxi_2,\bzeta_1,\bzeta_2 , \frac{z}{\eps} \Big)
 \exp \Big( \frac{i  z}{k_o \eps} (\bxi_2 \cdot \bzeta_2  +   \bxi_1 \cdot \bzeta_1) \Big) .
\end{equation}
Our goal is to study the asymptotic behavior of $\tilde{\mu}^\eps_{\btheta\btheta'}$ as $\eps \to 0$.
We have the following result, which shows that $\tilde{\mu}^\eps_{\btheta\btheta'}$ exhibits a multi-scale behavior
as $\eps \to 0$, with some components evolving at the scale $\eps$ and 
some components evolving at the order one scale. The proof is   similar to  the one
 of Proposition 1 in  \cite{garniers4},
but with a general incident field profile instead of a Gaussian one.

\begin{proposition}
\label{prop:sci1}%
If $\gamma_o\in L^1(\RR^d)$ and $\gamma_o({\bf 0})<\infty$, then
the function $\tilde{\mu}^\eps_{\btheta\btheta'}(\bxi_1,\bxi_2, \bzeta_1,\bzeta_2,z ) $ for $z \geq L$
can be expanded as
\begin{eqnarray}
\nonumber
&&
 \tilde{\mu}^\eps_{\btheta\btheta'}(\bxi_1,\bxi_2,  \bzeta_1,\bzeta_2,z)  =
\frac{K(L)}{\eps^{4d}}
\hat{f}  \Big( \frac{\bxi_1+\bxi_2+\bzeta_1+\bzeta_2}{2 \eps}-k_o\btheta \Big)
\overline{\hat{f}} \Big( \frac{\bxi_1+\bxi_2-\bzeta_1-\bzeta_2}{2 \eps} -k_o\btheta \Big)\\
\nonumber
&& \hspace*{1.in}\times
\hat{f}  \Big( \frac{\bxi_1-\bxi_2+\bzeta_1-\bzeta_2}{2 \eps}-k_o\btheta '\Big)
\overline{\hat{f}} \Big( \frac{\bxi_1-\bxi_2-\bzeta_1+\bzeta_2}{2 \eps}-k_o\btheta '\Big)
 \\
\nonumber
&& 
\quad
+
\frac{K(L)}{\eps^{3d}} 
\hat{f}  \Big( \frac{\bxi_1-\bxi_2+\bzeta_1-\bzeta_2}{2 \eps}-k_o\btheta'\Big)
\overline{\hat{f}} \Big( \frac{\bxi_1-\bxi_2-\bzeta_1+\bzeta_2}{2 \eps}-k_o\btheta'\Big)\\
\nonumber
&& \hspace*{1.in}\times
\hat{f}_1 \Big(\frac{\bzeta_2+\bzeta_1}{\eps}\Big)
A\big(\frac{\bxi_2+\bxi_1}{2} ,\frac{\bzeta_2 + \bzeta_1}{\eps} ,L \big) \\
\nonumber
&&  
\quad
+
\frac{K(L)}{\eps^{3d}} 
\hat{f}  \Big( \frac{\bxi_1+\bxi_2+\bzeta_1+\bzeta_2}{2 \eps} -k_o\btheta \Big)
\overline{\hat{f}} \Big( \frac{\bxi_1+\bxi_2-\bzeta_1-\bzeta_2}{2 \eps}-k_o\btheta \Big)\\
\nonumber
&& \hspace*{1.in}\times
\overline{\hat{f}_1} \Big(\frac{\bzeta_2-\bzeta_1}{\eps}\Big)
A \big(\frac{\bxi_2-\bxi_1}{2} ,\frac{\bzeta_2- \bzeta_1}{\eps} ,L \big) \\
\nonumber
&&  
\quad
+
\frac{K(L)}{\eps^{3d}} 
\hat{f}  \Big( \frac{\bxi_1-\bzeta_2+\bzeta_1-\bxi_2}{2 \eps}-k_o\btheta'\Big)
\overline{\hat{f}} \Big( \frac{\bxi_1-\bzeta_2-\bzeta_1+\bxi_2}{2 \eps}-k_o\btheta \Big)\\
\nonumber
&& \hspace*{1.in}\times
\hat{f}_{1} \Big(\frac{\bxi_2+\bzeta_1}{\eps} -k_o(\btheta'-\btheta)  \Big)
A\big( \frac{\bzeta_2+\bxi_1}{2} ,\frac{\bxi_2+ \bzeta_1}{\eps} ,L \big) \\
\nonumber
&& 
\quad
+
\frac{K(L)}{\eps^{3d}} 
\hat{f}  \Big( \frac{\bxi_1+\bxi_2+\bzeta_1+\bzeta_2}{2 \eps} - k_o \btheta \Big)
\overline{\hat{f}} \Big( \frac{\bxi_1-\bxi_2-\bzeta_1+\bzeta_2}{2 \eps} - k_o \btheta' \Big)\\
\nonumber
&& \hspace*{1.in}\times
\overline{\hat{f}_1} \Big(\frac{\bxi_2-\bzeta_1}{\eps}-k_o(\btheta'-\btheta) \Big)
A\big(  \frac{\bzeta_2-\bxi_1}{2} ,\frac{\bxi_2- \bzeta_1}{\eps}  ,L \big) \\
\nonumber
&& 
\quad
+\frac{K(L)}{\eps^{2d}} 
\hat{f}_1 \Big( \frac{\bzeta_2+\bzeta_1}{\eps}   \Big)
\overline{\hat{f}_1 }\Big( \frac{\bzeta_2-\bzeta_1}{\eps} \Big) \\
\nonumber
&& \hspace*{1.in}\times
A \big( \frac{\bxi_2+\bxi_1}{2},   \frac{\bzeta_2+ \bzeta_1}{\eps} ,L \big)
A \big( \frac{\bxi_2-\bxi_1}{2},   \frac{\bzeta_2- \bzeta_1}{\eps} ,L \big) \\
\nonumber
&&
\quad
 + \frac{K(L)}{\eps^{2d}} 
 \hat{f}_1 \Big( \frac{\bxi_2+\bzeta_1}{\eps}+ k_o(\btheta'-\btheta)\Big)
\overline{ \hat{f}_1 } \Big( \frac{\bxi_2-\bzeta_1}{\eps}+ k_o(\btheta'-\btheta)\Big) \\
\nonumber
&& \hspace*{1.in}\times
A \big( \frac{\bzeta_2+\bxi_1}{2},  \frac{\bxi_2+ \bzeta_1}{\eps} ,L  \big)
A \big( \frac{\bzeta_2-\bxi_1}{2},  \frac{\bxi_2- \bzeta_1}{\eps} ,L\big)
\\
&& 
\quad
 + R^\eps  ( \bxi_1,\bxi_2 ,  \bzeta_1 ,\bzeta_2 ,L)   ,
\label{eq:propsci11}
\end{eqnarray}
where the functions $K$ and $A$  are defined by
\begin{align}
\label{def:K}
K(z)  :=& \, 
 \exp\Big(- \frac{k_o^2}{2} \gamma_o({\bf 0}) z\Big) , \\
\nonumber
A(\bxi,\bzeta,z)   :=& \,
 \frac{1}{(2\pi)^d}
 \int  \Big[  \exp \Big( \frac{k_o^2}{4} \int_0^z \gamma_o \big( \bx + \frac{ \bzeta}{k_o} z' \big) \d z' \Big) -1\Big]\\
 & \times
   \exp \big( -i \bxi\cdot \bx  \big)
 \d\bx  ,   
\label{def:A}
\end{align}
the function $\hat{f}_1$ is
\begin{equation}
\hat{f}_1(\bzeta) = \int_{\RR^d} \hat{f} \big(\bk + \frac{\bzeta}{2}\big)
\overline{ \hat{f} }  \big(\bk - \frac{\bzeta}{2}\big)  \d \bk,
\end{equation}
and the function $R^\eps $ satisfies
$$
  \| R^\eps (\cdot,\cdot,\cdot,\cdot, L ) \|_{L^1(\RR^d\times \RR^d\times \RR^d\times \RR^d)} 
\stackrel{\eps \to 0}{\longrightarrow}  0  .
$$
\end{proposition}
Note that all terms in the expansion (except the remainder $R^\eps$) have $L^1$-norms of order one
when $\eps \to 0$.

\subsection{The intensity covariance function}\label{sec:scint2}
In the scintillation regime, we find from (\ref{eq:meanfluo2}) that the mean intensity is
\begin{eqnarray}
\EE\big[{\cal E}_\btheta]
=
\frac{1}{(2\pi)^{2d}}  \int_{\RR^d}  \overline{\hat{f}_1(\bzeta)} \exp\Big( \frac{k_o^2}{4} \int_0^L \gamma_o(-\bzeta \frac{z+L_0}{k_o}) -\gamma_o({\bf 0}) \d z \Big) \d\bzeta  \,
\hat{U}({\bf 0}) ,
\end{eqnarray}
which is independent of $\btheta$.
We get from (\ref{eq:covfluo2}), (\ref{eq:renormhatM2}), and Proposition \ref{prop:sci1} that 
the covariance function is of the form
\begin{align}
\nonumber
&{\rm Cov}({\cal E}_\btheta,{\cal E}_{\btheta'}) \\
\nonumber
&= 
\frac{1}{(2\pi)^{5d}}
\iint \Big| \int_{\RR^d}   \hat{f}_1\big(\bzeta+k_o(\btheta'-\btheta)\big)
\exp\Big( \frac{k_o^2}{4} \int_0^L \gamma_o(\bx -\bzeta \frac{z+L_0}{k_o})  \d z \Big) \d\bzeta \Big|^2\\
\nonumber
&\quad\quad \times 
e^{-i \bx \cdot \bxi } |\hat{U}(\bxi)|^2 \d\bxi  \d\bx  \exp\Big( - \frac{k_o^2\gamma_o({\bf 0}) L}{2}  \Big) \\
&\quad-
\frac{1}{(2\pi)^{4d}}
 \Big| \int_{\RR^d}  \hat{f}_1\big(\bzeta \big) \d\bzeta
 \Big|^2 |\hat{U}({\bf 0})|^2   \exp\Big( - \frac{k_o^2\gamma_o({\bf 0}) L}{2}  \Big)  ,
\end{align}
which is a function of $\btheta'-\btheta$ only.
Accordingly, we denote
\begin{equation}
\label{def:covintstat}
{\cal C} ( \btheta) :={\rm Cov}({\cal E}_{{\bf 0}},{\cal E}_\btheta) ={\rm Cov}({\cal E}_{\btheta'},{\cal E}_{\btheta'+\btheta})  . 
\end{equation}
If we assume that the source is a plane wave, so that $f(\bx)=1$ and $\hat{f}_1(\bzeta) = (2\pi)^{2d} \delta(\bzeta)$,
then we get 
\begin{equation}
\EE\big[{\cal E}_\btheta]
=
 \hat{U}({\bf 0})  =\int_{\RR^d} U(\bx) \d\bx ,
 \label{eq:mean1}
\end{equation}
and
\begin{align}
\nonumber
{\cal C}(\btheta)  =& \,
\frac{1}{(2\pi)^{d}}
\iint  
\exp\Big( \frac{k_o^2}{2} \int_0^L \gamma_o\big(\bx+  \btheta  (z+L_0)\big)  \d z \Big) e^{-i \bx \cdot \bxi } |\hat{U}(\bxi)|^2 \d\bxi  \d\bx \\
\nonumber
&\quad \times 
  \exp\Big( - \frac{k_o^2\gamma_o({\bf 0}) L}{2}  \Big) \\
&-
  |\hat{U}({\bf 0})|^2   \exp\Big( - \frac{k_o^2\gamma_o({\bf 0}) L}{2}  \Big)  .
  \label{eq:cov1}
\end{align}
In fact, Eqs.~(\ref{eq:mean1}) and (\ref{eq:cov1}) hold not only for plane waves, but for any illumination with uniform intensity
because $\hat{f}_1(\bzeta) = (2\pi)^{2d} \delta(\bzeta)$ if and only if $|f(\bx)|=1$ for all $\bx$.

The expression (\ref{eq:cov1}) of the intensity covariance function ${\cal C}(\btheta)$ is valid whatever the value of $L/\ell_{\rm sca}$.
In the strongly scattering regime, ie when the thickness of the scattering medium is larger than the scattering 
mean free path $L/\ell_{\rm sca} \gg 1$,
and if the function $\gamma_o$ can be expanded as (\ref{eq:expandgamma0}),
 then the  intensity covariance function ${\cal C}(\btheta)$ is
 \begin{align}
\nonumber
{\cal C}(\btheta) =& \,
\frac{1}{(\pi k_o^2 \bar{\gamma}_2 L)^{d/2}}
\int_{\RR^d}  |\hat{U}(\bxi)|^2 
\exp\Big( i \bxi \cdot  \btheta (\frac{L}{2}+L_0) \Big)\\
&\times \exp\Big( - \frac{|\bxi|^2}{k_o^2 \bar{\gamma}_2 L}
-\frac{k_o^2\bar{\gamma}_2 L^3}{48} | \btheta|^2\Big)
\d\bxi   .
\label{eq:covE3a}
\end{align}
We introduce $\ell_0$ the distance from the center of the scattering medium to the mask:
\begin{equation}
\label{def:ell0}
\ell_0 = L_0+\frac{L}{2}.
\end{equation}
It is shown in \cite[Proposition 6.3]{garniers3} that 
\begin{equation}
\label{def:rhoL}
\rho_L := \frac{2}{ \sqrt{k_o^2 \bar{\gamma}_2 L}}
\end{equation}
is the correlation radius of the speckle pattern transmitted at $z=L$.
In other words, the mask is illuminated by a speckle pattern with speckle radius $\rho_L$.
We therefore anticipate that the resolution of the image cannot be better than $\rho_L$,
and in fact it will turn out to be given by this length.

Using (\ref{def:ell0}) and (\ref{def:rhoL}) we can rewrite (\ref{eq:covE3a})
as
 \begin{equation}
{\cal C}(\btheta) = 
\frac{\rho_L^d}{(4\pi)^{d/2}}
\int_{\RR^d}   |\hat{U}(\bxi)|^2 
\exp\big( i \bxi \cdot  \btheta  \ell_0 \big)  \exp\Big( - \frac{\rho_L^2 |\bxi|^2}{4}
-\frac{L^2}{12 \rho_L^2} | \btheta|^2\Big)
\d\bxi   .
\label{eq:covE3}
\end{equation}
This shows that the intensity covariance ${\cal C}(\btheta)$ 
has the form of a peak centered at $\btheta={\bf 0}$ whose amplitude is
 \begin{eqnarray}
{\cal C}({\bf 0}) ={\rm Var}({\cal E}_\btheta) = \frac{\rho_L^d}{(4\pi)^{d/2}}
\int_{\RR^d}   |\hat{U}(\bxi)|^2 
 \exp\Big( -  \frac{\rho_L^2 |\bxi|^2}{4}\Big)
\d\bxi   .
\label{eq:covE3var}
\end{eqnarray}
The width $\Delta \theta_L$ 
of this peak is determined by the Gaussian in $\btheta$ and also by the Fourier transform in $\bxi$ in (\ref{eq:covE3}).
If the radius of the mask is of the order of $R_U$, then its Fourier transform is of radius of the order of $1/R_U$ and therefore
the width of the covariance function is of the order of
\begin{equation}\label{eq:width}
\Delta \theta_L= \frac{\sqrt{R_U^2+\rho_L^2}}{\ell_0} .
\end{equation}


\subsection{Extraction of the mask indicator function}\label{sec:scint3}
If we measure ${\cal E}_\btheta$ for a quasi-continuum of $\btheta$ in the domain $\Theta$,
then we can extract the empirical intensity covariance function
\begin{equation}
\label{eq:cC}
{\cal C}_{\rm emp}( \btheta) = \left<  {\cal E}_{\btheta'} {\cal E}_{\btheta'+\btheta} \right>_{\btheta'} - \left< {\cal E}_{\btheta'} \right>_{\btheta'}^2 ,
\end{equation}
where $ \left<  \cdot \right>_{\btheta'}$ stands for an average in $\btheta'$:
$$
 \left<  F(\btheta') \right>_{\btheta'} = \frac{1}{|\Theta|} \int_{\Theta} F(\btheta')\d \btheta'.
$$
Note that the term $\left<  {\cal E}_{\btheta'} {\cal E}_{\btheta'+\btheta} \right>_{\btheta'}$ has the form of an intensity autocorrelation.
Self-averaging (i.e. the fact that the empirical average is equal to the statistical expectation)
 is ensured provided the average in $\btheta'$ is carried out over a domain $\Theta$ that is large enough, ie 
a domain whose diameter $D_\Theta$ is larger than $\Delta \theta_L$
so that many speckle spots in $\btheta$ are integrated.
Then we have 
\begin{equation}
{\cal C}_{\rm emp}( \btheta) ={\cal C} ( \btheta)  ,
\end{equation}
where ${\cal C} ( \btheta) $ is the statistical intensity covariance function introduced in (\ref{def:covintstat}).
The self-averaging is efficient if the peak in the second moment of the intensity $\left<  {\cal E}_{\btheta'} {\cal E}_{\btheta'+\btheta} \right>_{\btheta'}$
emerges from the background 
and its fluctuations. The background is  the square mean intensity 
$\EE[{\cal E}_\btheta]^2= |\hat{U}({\bf 0})|^2 $,
and its fluctuations are of order $ |\hat{U}({\bf 0})|^2 / \sqrt{M}$ where $M$ is the number of speckle spots
recorded in the domain $\Theta$, which is  of the order of $(D_\Theta / \Delta \theta_L)^d$.
The amplitude of the main peak is the mean square intensity 
$\EE[{\cal E}_\btheta^2] = {\rm Var}( {\cal E}_\btheta )+|\hat{U}({\bf 0})|^2 $,
with ${\rm Var}( {\cal E}_\btheta ) \simeq [\rho_L^2/(R_U^2+\rho_L^2)]^{d/2} |\hat{U}({\bf 0})|^2$ by (\ref{eq:covE3var}).
Therefore the main peak can be extracted if
$
{\rm Var}( {\cal E}_\btheta ) \gg  |\hat{U}({\bf 0})|^2 / \sqrt{M}
$,
or equivalently,
$
\rho_L^2/(R_U^2+\rho_L^2) \gg { \Delta \theta_L}/ {D_\Theta}
$, that is to say, if
\begin{equation}
\label{eq:condDtheta}
D_\Theta \gg \frac{(R_U^2+\rho_L^2)^{3/2}}{\rho_L^2 \ell_0} .
\end{equation}
This condition means that we cannot image masks that are too large. 
\rc{If the mask is large then there are
 many speckle spots (with radius $\rho_L$) within the support of the mask indicator function and the 
 peak in the transmitted covariance function
 relative to the background, that is  ${\rm Var}( {\cal E}_\btheta )/|\hat{U}({\bf 0})|^2 \simeq [\rho_L^2/(R_U^2+\rho_L^2)]^{d/2}$, 
 becomes small;
moreover, the width of  the transmitted covariance function,
that is $\Delta \theta_L$ in Eq.~(\ref{eq:width}), 
becomes large, so the fluctuations of the background becomes large.  
Thus,  with a large mask the averaging in Eq.~(\ref{eq:cC})  becomes less efficient and the signal-to-noise ratio becomes low.}
 
From (\ref{eq:covE3}), we find that the Fourier transform of ${\cal C}(\btheta)$, which is 
the power spectral density of the stationary random process ${\cal E}_\btheta$ by Bochner's theorem, which  is nonnegative,  
moreover, 
 is of the form
\begin{align}
\nonumber
\hat{\cal C} (\bpsi) :=&\, \int_{\RR^d} {\cal C}(\btheta) e^{i \bpsi\cdot\btheta} \d\btheta \\
=&\,
\frac{(12)^{d/2} \rho_L^d}{L^d}
\int_{\RR^d} |\hat{U}(\bxi)|^2 \exp\Big( - \frac{\rho_L^2 |\bxi|^2}{4}\Big) 
\exp \Big( - \frac{ 3 \rho_L^2 \ell_0^2}{L^2} \big| \bxi - \frac{\bpsi}{\ell_0}\big|^2\Big)\d\bxi  .
\end{align}
If we introduce 
\begin{equation}
{R_L}:= \frac{\sqrt{3} \ell_0 \rho_L}{L} ,
\end{equation}
then the observed power spectral density can be written as
\begin{equation}
\hat{\cal C} \big( \ell_0 \bk \big)   
=
\frac{(12)^{d/2} \rho_L^d}{L^d}
\int_{\RR^d}  |\hat{U}(\bxi)|^2 \exp\Big( - \frac{\rho_L^2 |\bxi|^2}{4}\Big) 
\exp \big( - R_L^2 | \bxi - \bk|^2 \big)\d\bxi .
\end{equation}

If we know the thickness of the scattering medium $L$ and the distance $\ell_0$,
then this shows that we can extract $|\hat{U}(\bxi)|$
for spatial wavevectors $\bxi$ with modulus smaller than $2 /\rho_L$, and on a grid of wavevectors 
with grid step $1/R_L$.
Indeed  the damping by the exponential  
\[ \exp\Big( - \frac{\rho_L^2 |\bxi|^2}{4}\Big) ,
\]
 limits the resolution to the scale of
the speckle pattern,  the scale $\rho_L$, while the convolution by the exponential 
\[
\exp \big( - R_L^2 | \bxi |^2 \big) ,
\]
 limits the  maximum radius of the object 
that can be imaged 
to $R_L$ due to the finite range of the memory effect.
If we do not know these distances, then we can still extract $|\hat{U}(\bxi)|$ on a grid 
that is known up to multiplicative factor. As a consequence, we will be able to get the shape of the mask,
but up to a dilation.
However, a very interesting point is that one can get an image of the mask whatever its range.

If the mask is close to the scattering medium $L_0=0$, or even embedded in it, then we can hardly get any image
at all as $1/R_L = 2/(\sqrt{3}\rho_L)$.
If the mask is far from the scattering medium (far in the sense that the distance $L_0$
from the scattering medium to the mask is larger than the thickness $L$ of the scattering medium),
then the grid of wavevectors is rich (it contains $(2R_L/\rho_L)^d\simeq (2\sqrt{3} L_0/L)^d$ wavevectors)
and we can get detailed information on  $|\hat{U}(\bxi)|$ for $|\bxi|$ smaller than $2 /\rho_L$.

Finally, since we know the phase of $U(\bx)$ (it is a non-negative function so the phase is zero),
it is possible to apply a phase-retrieval algorithm to extract the profile of $U(\bx)$
from the modulus of its Fourier transform, for instance using a Gerchberg-Saxon-type iterative algorithm \cite{fienup}.
The resolution of the extraction will be of the order of $\rho_L$.
In other words, we can image the mask with a resolution of the order of $\rho_L$.
The mask should not be too large so as to satisfy (\ref{eq:condDtheta}):
$R_U^3 \ll D_\Theta \rho_L^2 \ell_0$.

\section{Summary of the main results}
\label{sec:summ}

We summarize the steps taken to get the image of the mask.   
First recall the configuration in Figure \ref{fig:1}. 
A  time-harmonic source emits a probing quasi plane-wave
 coming from the left and hits a  mask
 after propagating through the medium.
 Then the total transmitted intensity  is 
 measured, that is
 \begin{equation}
\label{def:Etheta2-sum}
{\cal E}_\btheta = \int_{\RR^d} |E_\btheta(\bx,L+L_0) |^2 U(\bx) \d\bx  ,
\end{equation}
for $E_\btheta(\bx,L+L_0)$ the probing time-harmonic field at the mask
and $\btheta$ the incident angle. 
On the one hand, we form the empirical intensity covariance function ${\cal C}_{\rm emp} (\btheta)$ by 
evaluating the autocorrelation in (\ref{eq:cC}).
On the other hand, we model the complex or random section as a random medium.
The statistical intensity covariance function is then of the form
\begin{align}
\nonumber
{\rm Cov}( {\cal E}_{\btheta},{\cal E}_{\btheta'}) =& \iint
 \EE \big[ |E_\btheta(\bx,L+L_0) |^2 |E_{\btheta'}(\bx',L+L_0) |^2 \big] U(\bx) U(\bx') \d\bx \d\bx'  \\
&  - 
\EE [ {\cal E}_\btheta]\EE [ {\cal E}_{\btheta'}]  .
\label{eq:cCs-sum}
\end{align}
We moreover consider a propagation regime corresponding to paraxial waves,
that is, we assume that the wavelength associated with the probing frequency
is small compared to the correlation radius of the medium,  which in turn
is small compared to the thickness of the random section. Furthermore, we consider
the scintillation regime associated with the paraxial wave equation corresponding 
to an incoming quasi plane-wave.  Then the statistical covariance function
for incident angles $\btheta, \btheta'$ is
\begin{align}
\nonumber
{\rm Cov}({\cal E}_\btheta,{\cal E}_{\btheta'}) =& \,
\frac{1}{(2\pi)^{d}}
\iint  
\exp\Big( \frac{k_o^2}{2} \int_0^L \gamma_o\big(\bx+  (\btheta'-\btheta) (z+L_0)\big)  \d z \Big) e^{-i \bx \cdot \bxi } |\hat{U}(\bxi)|^2 \d\bxi  \d\bx \\
\nonumber
 &\quad \times 
  \exp\Big( - \frac{k_o^2\gamma_o({\bf 0}) L}{2}  \Big) \\
 &-
  |\hat{U}({\bf 0})|^2   \exp\Big( - \frac{k_o^2\gamma_o({\bf 0}) L}{2}  \Big)  .
  \label{eq:cov1-sum}
\end{align}
Here, $L$ is the thickness of the random section, $L_0$ is the 
distance from the random section to the hidden mask,
$k_o=\omega/c_o$ is the wavenumber for $c_o$ the background 
speed,  $\gamma_o$ is the lateral covariance function of the 
medium fluctuations  (the medium covariance function integrated with 
respect to the range coordinate, see Eq.~(\ref{def:gamma0:chap10}))
and $U$ is the mask with Fourier transform~$\hat{U}$. 

It turns out that the statistical intensity covariance function depends  only the 
angle difference    $\btheta'-\btheta$, and the mean intensity is constant.
This means that we can view ${\cal E}_\btheta$  as a wide sense stationary field in $\btheta$
and we can denote 
$$
{\cal C}(\btheta) = {\rm Cov}({\cal E}_{{\bf 0}},{\cal E}_\btheta)=
{\rm Cov}({\cal E}_{\btheta'},{\cal E}_{\btheta'+\btheta}).
$$
 
Several questions then need to be answered: (i) how close is 
the empirical covariance function ${\cal C}_{\rm emp}(\btheta)$  to the statistical covariance function ${\cal C}(\btheta) $; (ii) how can we 
unravel the shape of the mask  $U$ from the empirical  covariance
and with what resolution?  

We can answer these questions when 
the medium fluctuations are smooth so that $\gamma_o$ also is smooth,
moreover, when the random section is relatively thick so that 
 $  k_o^2\gamma_o({\bf 0}) L \gg 1$. 
 Albeit, corresponding to stronger assumptions, this context allows us
 to get explicit and simple answers to the above questions. 
 Note first that the thickness of the random section $L$ is large
 compared to the scattering mean free path, so that the coherent field,
 corresponding  to  the second term in the right-hand side of Eq. (\ref{eq:cov1-sum}),   can be ignored as
 it is vanishingly small.  We then find that the statistical intensity covariance function ${\cal C}(\btheta)$  becomes 
\begin{equation}
{\cal C}(\btheta) 
=
\frac{\rho_L^d}{(4\pi)^{d/2}}
\int_{\RR^d}   |\hat{U}(\bxi)|^2 
\exp\big( i \bxi \cdot  \btheta  \ell_0 \big)  \exp\Big( - \frac{\rho_L^2 |\bxi|^2}{4}
-\frac{L^2}{12 \rho_L^2} | \btheta|^2\Big)
\d\bxi   ,
\end{equation}
where $\rho_L $ is the correlation radius of the speckle pattern  at depth $z=L$:
  \begin{equation}
\label{def:rhoL2-sum}
\rho_L := \frac{2}{ \sqrt{k_o^2 \bar{\gamma}_2 L}} ,
\end{equation}
 and  $\bar{\gamma}_2$ is
  defined in Eq. (\ref{eq:expandgamma0}) and is a measure
  of the strength of the medium fluctuations. 
 We can write this  as
  \begin{equation}
{\cal C}(\btheta)  =
 \left(\left( U  \star U \right) * {\mathcal S}_{\rho_L/\sqrt{2}} \right)( \btheta\ell_o) 
\exp\Big( -\frac{L^2}{12 \rho_L^2} |\btheta|^2\Big)  ,
\label{eq:int-sum}
\end{equation}
where ${\mathcal S}_{\sigma}$ is the Gaussian density with mean zero and standard 
deviation $\sigma$, 
the symbol $*$ denotes the convolution operation, and
the symbol $\star$  denotes the autocorrelation operation:
$$
f*g(\btheta) = \int f(\btheta')g(\btheta-\btheta') {\rm d} \btheta', \quad \quad
f\star f(\btheta) = \int f(\btheta')f(\btheta+\btheta') {\rm d} \btheta' .
$$
Note that the convolution with the Gaussian gives a smoothing of the
data on the resolution scale $\rho_L$  while the damping
by the last exponential term means that the size of the mask cannot be too
large because the memory effect is then not valid. Explicitly 
this last condition means that we must have 
  \begin{equation}
    R_U   <  \rho_L \left( 1 + 2 \frac{ L_0}{L} \right)  ,
 \end{equation}
for $R_U$ the radius of the mask.  We see, therefore, that for the scheme 
to work well we need the distance from the random section to the
mask to be large relative to the thickness of the random section.
Consequently, with the thickness of the random section being fixed, it is better 
that the random section is located closer to the 
source. Then the scale of the mask that can be imaged within 
the memory angular aperture is larger. 
This is indeed a type of shower curtain effect \cite{ishimaru}.

We remark moreover that according to the physical arguments discussed in \cite{bertolotti,katz14}
we should have  for the recorded total intensity  
as a function of incident angle
$$
{\cal E}_\btheta\simeq  \left( U  * S \right)(\btheta)   ,
$$
up to a dilation and scaling,
where
 the speckle pattern $S$ is {\it common} for the different 
 probing angles $\btheta$ due to the memory effect.  
 We then get for the empirical intensity autocorrelation function  
$$
{\cal E} \star {\cal E} =
    \left( U  * S \right) \star \left( U  *  S \right)   = 
    \left( U  \star U  \right) *   \left( S  \star S \right)  \simeq
   \left( U  \star U  \right) *  \EE  \left( S  \star S \right) ,
$$
where, in the last approximation, we mean that the relative
  fluctuations in the autocorrelation of the speckle pattern $S$ is  small
because its correlation radius is small.  
  We see  that this conforms with the result in Eq.~(\ref{eq:int-sum}) 
  for  ${\mathcal S}_{\rho_L/2}$  replaced by $\EE  \left( S  \star S \right)$.

 The question (i) posed above still remains. To what extent 
will the empirical covariance function of the total intensity, that
is the covariance estimated from real data, be a good 
approximation for   
$ {\cal C}(\btheta )$~?
In the regime of deep probing we obtain the  empirical  covariance function
via averaging over angles
as in Eq. (\ref{eq:cC}).
 For statistical stability we need that this averaging takes place
 over many speckle spots and for this  we need the condition in 
 Eq.  (\ref{eq:condDtheta})  to be fulfilled, which means that the size of the mask cannot be too large for statistical stability.
   In \cite{bertolotti} it was  numerically checked that the signal-to-noise ratio 
   is proportional to the square root of the number of speckle spots included in the scanned range.  
   Such a behavior  regarding the signal-to-noise ratio  was also observed empirically in
\cite{roy}.
  We remark also that in \cite{bertolotti}
    the authors  found it   experimentally  convenient, in view of long scan times, 
    to make a number of non-overlapping scans and average their covariances rather than taking one large scan. 
  
Finally, in order to recover $U$ from an estimate of $U \star U$ one can apply
a phase retrieval algorithm as discussed in Section \ref{sec:scint3}.
Note moreover   that in view of  Eq. (\ref{eq:int-sum})  we find that  if $\ell_0$ is known we can estimate
the shape of the mask while if this parameter is unknown we can estimate the shape 
only up to a dilation. 

\section{Conclusions}
\label{sec:final}
 
We  have considered imaging of a mask hidden behind
a scattering medium via forming the covariance function of the total transmitted
 intensity with respect to the incident angle of the probing beam.
This procedure serves to stabilize the effect of the random medium. 
 The physical picture is that,
 due to the  memory effect, the speckle pattern illuminating the mask
 is approximately the same one, but it is shifted for different incident angles, within the memory range.  
 The empirical covariance function is then essentially the autocorrelation
 of the mask convolved with the autocorrelation 
 of this approximately invariant speckle pattern.  
 The autocorrelation function of this speckle pattern  is a sharply peaked 
 statistically stable function because its correlation radius is small.
 
 Our mathematical analysis  gives a more general
 description of the empirical covariance function than the one just described
 based on physical arguments.
 We obtain a corrected expression for the intensity covariance function
  and explicit expressions for the resolution and the signal-to-noise ratio.
  We explain how they relate to the  configurational parameters and we clarify  the role of the memory effect.
We remark that, in the appropriate limit, explicitly  articulated  here,  the above description
 based on physical arguments gives the correct picture.
 
We show that the resolution or blurring scale associated with 
 the convolution by  the autocorrelation of the speckle pattern
 is the correlation radius of the speckle pattern.  
 We give an explicit condition regarding the sampling
 over incident angle for the empirical covariance function to be statistically
 stable and have a high signal-to-noise ratio; this condition means that the mask cannot be too large.  
 The size of the mask that can be imaged is also limited  by the  memory effect.
 We give an explicit
 bound on the size of the mask so that one stays within the memory
 effect of the configuration. 
   
As a result, the empirical intensity covariance function over incident angle is
 a (blurred) version of the autocorrelation of the mask.
 The mask itself can then be estimated via a phase retrieval
 algorithm. We do not consider this step here, however, this 
 step has been implemented in the various physical 
 experiments regarding speckle imaging.  
 
 We remark that the same analysis applies when 
 a  mask is wedged in between two random sections and 
 the total transmitted intensity is measured. \rc{This is because
 the total transmitted intensity  is measured and the medium is lossless,  so that
 the random section after the mask plays no role
 in either case.} 
  
 Recall  also that   our analytic expression for the covariance function
of the transmitted intensity in Eq. (\ref{def:Etheta2-sum}) is more general than the standard 
 one  derived from physical arguments which corresponds
 to the description in Eq.~(\ref{eq:int-sum}).  Thus, the more general expression,
 based on weaker assumptions,  could potentially form the basis
 for an iterative image enhancement procedure when one is not deep
 into the scintillation regime.
 Moreover, the more general  expression gives insight about for which experimental
 configuration the conventional  speckle imaging procedure works well, respectively not so well.  
 
 \rc{Finally, recall that we have considered the scintillation regime in which
 the incoming beam radius is large relative to the correlation radius of the scattering medium.
 In the  spot-dancing regime  the incoming beam radius  is small relative to the 
 correlation radius of the scattering medium, and the analysis shows that
 the beam is essentially subjected to a random displacement \cite{garniers3}, but
 does not form a speckle pattern which is the basis for the speckle imaging 
 approach. Thus the proposed approach for imaging does not work
 well in the spot-dancing regime. }

\section*{Acknowledgements}
  This research   it is supported in part by 
 AFOSR grant  FA9550-18-1-0217, NSF  grant 1616954, 
 Centre Cournot, Fondation Cournot, and 
 Universit\'e Paris Saclay (chaire D'Alembert).


\end{document}